\documentclass[twocolumns]{aa}
\usepackage{graphicx}
\usepackage{txfonts}
\usepackage{natbib}
\begin{document}
   \title{First detection of [CII]158$\mu$m at high redshift:\\
   vigorous star formation in the early universe}

   \subtitle{}

  \author{R.~Maiolino\inst{1}
          \and
          P.~Cox\inst{2}
          \and
          P.~Caselli\inst{1}
          \and
          A.~Beelen\inst{3}
          \and
          F.~Bertoldi\inst{4}
          \and
          C.~L.~Carilli\inst{5}
          \and
          M.~J.~Kaufman\inst{6}
          \and
	  K.~M.~Menten\inst{3}
	  \and
          T.~Nagao\inst{1,7}
          \and
          A.~Omont\inst{8}
          \and
          A.~Wei{\ss}\inst{9,3}
          \and
          C.~M.~Walmsley\inst{1}
          \and
	  F.~Walter\inst{10}
          }

   \offprints{R. Maiolino}

   \institute{INAF - Osservatorio Astrofisico di Arcetri,
      L.go E. Fermi 5, I-50125 Firenze, Italy
      \and
   IRAM, 300 Rue de la Piscine, F-38406, St-Marin-d'H\`eres, France
         \and
	 Max-Planck-Institut f\"ur Radioastronomie, Auf dem H\"ugel 69,
D-53121 Bonn, Germany
         \and
   Radioastronomisches Institut, Universit\"{a}t Bonn,
    auf dem H\"{u}gel 71, D-53121 Bonn, Germany
         \and
	 National Radio Astronomy Observatory, P.O. Box O,
Socorro, NM 87801, USA
	 \and
     Department of Physics, San Jose State University,
       1 Washington Square, San Jose, CA 95192, USA
         \and
     National Astronomical Observatory of Japan, 2-21-1 Osawa, 
Mitaka, Tokyo 181-8588, Japan
         \and
      Institut d'Astrophysique de Paris, UMR7095 CNRS, Universit\'e Pierre \& Marie Curie, 98 bis boulevard Arago,  F-75014 Paris , France
         \and
     Instituto de Radioastronomia Milimetrica, Avenida Divina Pastora 7,
     SP-18012 Granada, Spain
	 \and
	 Max-Planck-Institut f\"ur Astronomie, K\"onigstuhl 17,
     D-69117 Heidelberg, Germany}

   \date{Received ; accepted }

   \abstract{We report the detection of the
$^2P_{3/2} \rightarrow {^2P_{1/2}}$ fine-structure line of $\rm C^+$ at
157.74~$\rm \mu m$ in SDSSJ114816.64+525150.3 (hereafter J1148+5251),
the most distant known quasar, at $z=6.42$, using the IRAM 30-meter telescope.
This is the first detection of the [C{\sc II}] line at high
redshift, and also the first detection in a Hyperluminous Infrared
Galaxy ($\rm L_{FIR} > 10^{13} \, L_\odot$). The [C{\sc II}] line is
detected at a significance level of 8$\sigma$ and has a
luminosity of $\rm 4.4\times 10^9 \,L_\odot$. The $\rm L_{[CII]}/L_{FIR}$ ratio
is $\rm 2\times 10^{-4}$,
about an order of magnitude smaller than observed in local
normal galaxies and similar to the ratio observed in local Ultraluminous
Infrared Galaxies. The [C{\sc II}] line luminosity indicates that the host
galaxy of this quasar is undergoing an intense burst of star formation
with an estimated rate of $\rm \approx 3000 \, M_\odot \, yr^{-1}$.
The detection of
$\rm C^+$ in SDSS~J1148+5251 suggests a significant enrichment of metals
   at $z\sim 6$ (age of the universe $\sim$870~Myr), although
   the data are consistent with a reduced carbon to oxygen ratio as
   expected from chemical evolutionary models of the early phases
   of galaxy formation.

   \keywords{Galaxies: high redshift -- Galaxies: ISM --
   quasars: individual: J114816.64+525150.3 -- 
   Infrared: galaxies --  Submillimeter --
   ISM: abundances }
   }

   \maketitle
%

\section{Introduction}

The $^2P_{3/2} \rightarrow {^2P_{1/2}}$ fine-structure line of $\rm C^+$
at 157.74~$\rm \mu m$ is often the brightest emission line in the
spectrum of galaxies, accounting for as much as $\sim$0.1-1\% of
their total luminosity \citep[][]{crawford85,stacey91,wright91}.
The line is emitted predominantly by
gas exposed to ultraviolet radiation in
photo dissociation regions (PDRs) associated with star forming activity
(even in galaxies hosting AGNs),
and has been extensively used to investigate the physical conditions
of PDRs and to trace star formation in external galaxies
\citep[e.g.][]{malhotra01,luhman03,kaufman99,genzel00,boselli02,pierini03}.

In local galaxies, with far-infrared luminosities\footnote{We adopt
the $\rm L_{FIR}=L(40-500\mu m)$
definition given in \cite{sanders96}.} $\rm L_{FIR} < 10^{12}
\, L_\odot$,
the fractional contribution of the $\rm C^+$ luminosity
to the far-infrared luminosity, which
is a measure of the gas heating efficiency, is typically in the range $\rm -3
\le log (L_{[CII]}/L_{FIR}) \le -2$ (e.g., Stacey et al. 1991). For sources with
far-infrared luminosities in excess of $\rm 10^{12} \, L_\odot$ or
with increasing star formation activity, this ratio drops by an order
of magnitude \citep[][]{malhotra01,luhman98,luhman03,negishi01}.
Various possibilities have been proposed to explain this effect at
high luminosities, specifically:
1) a high ratio of ultraviolet flux to gas density, resulting
in positively charged dust grains which in turn reduce the efficiency
of the gas heating by the photoelectric effect \citep[e.g.][]{kaufman99};
2) opacity effects which weaken the [C{\sc II}] emission
line in infrared luminous galaxies
\citep[][]{gerin00,luhman98}; 3) contribution to $\rm L_{FIR}$ from
dust associated with HII regions \citep[][]{luhman03}; 4) contribution to
$\rm L_{FIR}$ from an AGN \citep[see][]{malhotra01}.

The possibility of probing the interstellar medium and tracing star
formation in galaxies at cosmological distances by using the bright
$\rm C^+$ emission line, when redshifted into submillimeter atmospheric
windows, was proposed by \cite{petrosian69}, and
further discussed by \cite{loeb93}, \cite{stark97} and
\cite{blain00}.  However, [C{\sc
II}] has so far only been detected in local galaxies ($z<0.1$)
from space and airborne observatories
\citep[e.g.][]{luhman03}, while all the searches for $\rm C^+$ at higher
redshifts have been unsuccessful. Deep searches for $\rm C^+$ were 
carried out in
$z>3$ infrared luminous galaxies and quasars having massive reservoirs
of molecular and neutral gas detected in CO emission lines. 
Specifically, $\rm C^+$ was searched for in
the $z=3.13$ damped Ly$\alpha$ absorption system towards
PC~1643+4631A \citep[][]{ivison98},
the $z=4.12$ quasar
PSS~2322+1944 (Benford et al., in prep.),
the $z=4.69$ quasar BR~1202$-$0725 \citep[][Benford et al. in prep.]{isaak94,werf99}, a $z=4.92$ lensed
galaxy in the cluster CL~1352+62 \citep[][]{marsden05}, and
the $z=6.42$ quasar
SDSS~J1148+5251 \citep[][the same object discussed here]{bolatto04}.

In this letter, we report the detection of $\rm C^+$ in
J114816.64+525150.3 (hereafter J1148+5251)
the most distant quasar known whose redshift of 6.42 puts it at the
end of the epoch of re-ionization \citep[][]{fan03}.
Millimetric and submillimetric continuum detections indicate that the host galaxy of
this quasar radiates a far-IR luminosity\footnote{Thanks to new submm data
Beelen et al. (in prep.) obtained a new estimate of the far-IR
luminosity, which is in the range $\rm 1.2-3.2 \times 10^{13}~L_{\odot}$.}
of $\rm \approx 2\times 10^{13}\, L_\odot$,
which, if ascribed to starburst activity (and not to the QSO itself),
would imply star formation rates of a few times $\rm 10^3~M_{\odot}~yr^{-1}$
\citep[][]{bertoldi03a,robson04}.
A massive reservoir of molecular and neutral gas
of a few $\rm 10^{10} \, M_\odot$ has been detected via CO line emission
\citep{walter03,walter04,bertoldi03b}.
The detection of copious amounts of dust and molecular gas indicates
that the insterstellar medium in J1148+5251 is significantly enriched in
metals. This enrichment is also seen through the detection of 
prominent iron line emission in the rest-frame UV spectrum of the quasar
\citep{maiolino03}, although the latter samples its nuclear region.

The deep observations described in this paper, improve significantly
the noise--level reported by \cite{bolatto04}
and enabled us to achieve the first detection of the $\rm C^+$ emission
line at cosmological distances as well as the first detection of this line in
a galaxy with an infrared luminosity in excess of $10^{13} \,
L_\odot$. We discuss the implications of this detection for the
physical properties of the ISM and of star formation in J1148+5251
\footnote{In this letter we assume the concordance $\Lambda$-cosmology with
$\rm H_0=71~km~s^{-1}Mpc^{-1}$, $\rm \Omega _{\Lambda}=0.73$ and
$\rm \Omega _{m}=0.27$ \citep[][]{spergel03}}.


\begin{table*}
\caption{Properties of the [CII] line observed toward SDSS J1148+5251
compared with the CO(6-5) line observed by \cite{bertoldi03b}.}
\label{obs}     
\centering      
\begin{tabular}{l c c c c c c c} 
\hline\hline                
Line  & $\rm \nu _{rest}$ & $\rm \nu _{obs}$ & $z_{\rm line}$ & Peak int. &
   $\rm \Delta V _{FWHM}$ & I & L \\
      & \multicolumn{2}{c}{[GHz]} &   & [mJy] & [$\rm km~s^{-1}$] &
       [$\rm Jy~km~s^{-1}$] & [$\rm 10^9~L_{\odot}$]\\
\hline                   
 [CII] ($\rm ^2P_{3/2}-^2P_{1/2}$) & 1900.54 & 256.172 & 6.4189$\pm$0.0006 &
      11.8 & 350$\pm$50 & 4.1$\pm$0.5 & 4.4$\pm$0.5 \\
 CO (6--5) &  692.473 & 93.204  & 6.4189$\pm$0.0006 & 2.45 & 279$\pm$65  &
   0.73$\pm$0.076  & 0.29$\pm$0.02 \\
\hline
\end{tabular}

\end{table*}
   \begin{figure}
   \centering
   \includegraphics[width=8.5truecm]{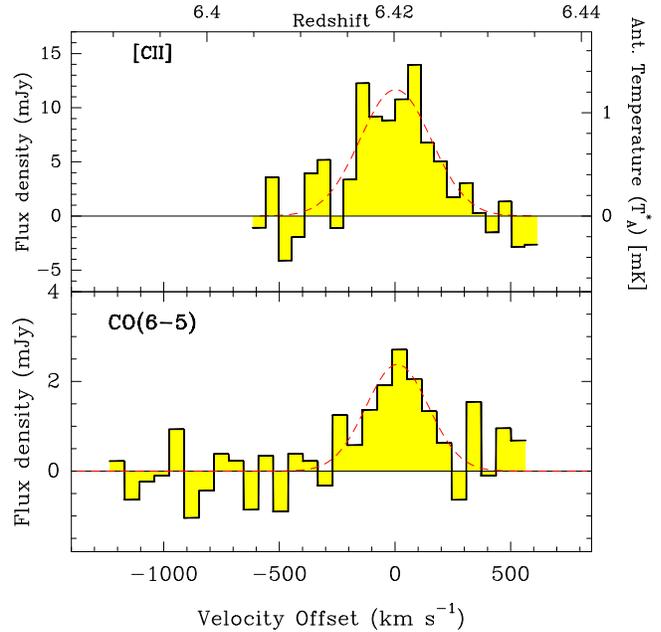}
      \caption{Spectrum of the [C{\sc II}] 157.74~$\rm \mu m$ emission line 
in the quasar J1148+5251 at $z=6.42$ shown with a velocity
resolution of $\rm 56 \, km \, s^{-1}$ (top panel) compared
to the CO(6--5) emission line (bottom panel - from
Bertoldi et al. 2003b). The dashed curves show the
gaussian fits to the line profiles (see Table~\ref{obs}).
              }
         \label{cii}
   \end{figure}

\section{Observations and results}

Observations were carried out with the IRAM 30m telescope
in two observing runs: February 2--3 and March 8--10, 2005.
We used alternatively the C and D 150 receivers with 
filter banks covering a bandwidth of 1~GHz
with 256 channels spaced by 4~MHz.
The receivers were tuned to 256.1753~GHz corresponding to the frequency
of the [C{\sc II}] emission line (rest frequency 1900.539~GHz)
redshifted to $z=6.4189$, the redshift of the CO emission lines,
but we also observed with the frequency offset by $\pm$100~km/s
to minimize, and check, the effects of possible instrumental artifacts.
At this frequency the Half Power Beam Width is 9.6$''$
and the 1GHz bandwidth corresponds to $\rm 1170 \, km \,s^{-1}$.
The observing conditions were generally good: low opacity, with
$\rm \tau _{256GHz}< 0.1$ during the March run and
 $\rm \tau _{256GHz}<0.2$ during the February run, and  
system temperature T$_{sys}\sim$350~K.
The observations were done in wobbler switching mode
with a switching frequency of 0.5~Hz and a wobbler throw of 200$''$.
In total, the source was observed for 12.4 hours.
Calibration was obtained every 15 minutes using the standard
hot/cold-load absorber measurements. Pointing was checked about every
1.5 hour by means of one of the two 3mm receivers (A100/B100) and
the pointing accuracy was found to be in the range $1-3^{\prime\prime}$.
The focus was checked about every 3--4 hours, and was found not
to change significantly.

The data were processed using the CLASS software.
After dropping some bad scans, only linear baselines were
subtracted from individual spectra. The individual scans
were realigned in frequency and co-added by weighting with 1/rms$^2$.
The resulting profiles were re-gridded to a velocity resolution
of 56~km/s leading to an rms of about 0.3~mK
(2.8~mJy)\footnote{The conversion between antenna temperature and
flux density scale at 256~GHz for the 30m is 9.5~Jy/K.}.
We estimate the flux density scale to be accurate to about 30\%.

The final spectrum of the [C{\sc II}] emission line in J1148+5251 is
shown in Fig.~\ref{cii}. The $\rm C^+$ fine-structure line is detected with a
confidence 8~$\sigma$. The velocity integrated flux is $\rm
0.44 \pm 0.05~K \, km \, s^{-1}$ (in antenna temperature,
$\rm T_{mb}=1.91~T_A$),
or $\rm 4.1 \pm 0.5 \, Jy \, km \,
s^{-1}$, and the line center corresponds within the uncertainties
to the center of the CO line emission (Table~\ref{obs} and Fig.~\ref{cii}).

\cite{bolatto04} used the
James Clerk Maxwell Telescope to detect the C$^+$ line in J1148+5251, but
could only place an upper limit to its line strength.
To compare their limits with our result we note that
the peak intensity of our line
(12~mJy) is three times lower than the rms (32~mJy) in their final
spectrum smoothed to a resolution of 10~MHz (or $\rm 12 \, km \, s^{-1}$).
The [C{\sc II}] velocity integrated 
flux reported in this paper is twice larger than the 1$\sigma$ limit estimated
by Bolatto et al. for the intensity of any putative line in the
JCMT spectrum.

   \begin{figure}
   \centering
   \includegraphics[width=8.5truecm]{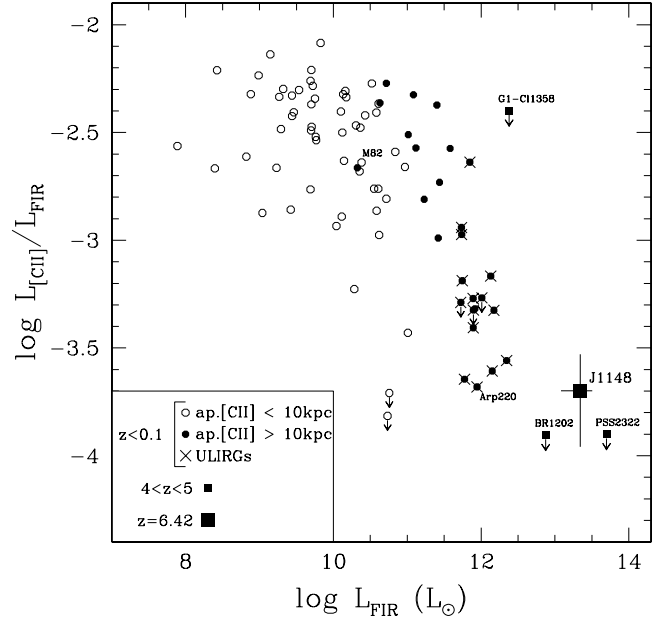}
      \caption{
      The $\rm L_{[CII]}/L_{FIR}$ ratio versus $\rm L_{FIR}$ for
normal and starburst local galaxies (circles), as measured by the ISO
Long Wavelength Spectrometer (LWS), and for high-$z$ sources
(squares), as measured from the ground.  J1148+5251 is shown with a
large square.
The small squares indicate the upper limits for the
three other high-$z$ sources (identified by their names) where [CII]
was searched for.  Empty circles indicate galaxies for which the
aperture 
      to measure [CII] (i.e. ISO--LWS) is smaller than 10~kpc, and therefore
      it may sample a smaller region relative to the IRAS beam used for
      the far-IR flux.
      Crosses indicate local ULIRGs ($\rm L_{IR}>10^{12}~L_{\odot}$).
      For reference, well-known 
	local galaxies (Arp220, M82) are also identified.}
         \label{cii_fir}
   \end{figure}

\section{Discussion}

The emission of $\rm C^+$ in J1148+5251 is an
extreme example of what is seen in local infrared galaxies.
The $\rm L_{[CII]}/L_{FIR}$ ratio in J1148+5251 is $2 \times 10^{-4}$, about
an order of magnitude smaller than in local starburst galaxies. This
is illustrated in Fig.~\ref{cii_fir}, where we plot the $\rm L_{[CII]}/L_{FIR}$
ratio in nearby star-forming galaxies as a function of their
far-infrared luminosity, together with the current upper limits for
high-redshift sources and the value obtained for J1148+5251. A
decrease of the $\rm L_{[CII]}/L_{FIR}$ ratio with increasing $\rm
L_{FIR}$ beyond $\rm 10^{11.5} \, L_\odot$ is clearly apparent
in Fig.~\ref{cii_fir}. 
The  $\rm L_{[CII]}/L_{FIR}$ ratio for J1148+5251 is consistent with 
this general trend.
Note also that our sensitive detection is
consistent with the upper limits inferred for other high-z QSOs in Fig.~\ref{cii_fir}.

The detection of strong [CII] emission suggests that the interstellar
medium was already significantly enriched with metals at $z=6.4$, i.e.
when the universe was as young as 870~Myr. However, the chemical evolutionary
models predict that the enrichment of carbon relative to the $\alpha$ elements
is delayed (since carbon is mostly produced by intermediate/low mass stars),
and in particular the C/O ratio should be about 1/6 in the young
stages of massive elliptical galaxies, while the absolute abundance
of O should be similar to the local value,
at least at an age of about 1~Gyr \citep[][]{pipino04}.
To achieve a better
physical understanding of this source, and in particular to
investigate whether the C-poor scenario is consistent with the data or not,
we have compared all available observational data for the host
galaxy of the quasar (i.e. [CII], FIR and CO transitions) with PDR models.
The relevant observational data are summarized in the second column of
Tab.\ref{mod}. We used the PDR model of \cite{kaufman99},
modified for the conditions
appropriate for J1148+5251.
In particular, we lowered the carbon-to-oxygen abundance
as discussed above.
We did not include detailed models for the evolution
of dust \citep[][]{morgan03}, nor the possible contribution of dust from
SNe \citep[][]{maiolino04}, we simply scaled the the abundance of dust
grains and PAHs proportionally to the carbon abundance.
The model requires high densities and high UV radiation fields to
reproduce the data. However, a PDR model which fits observations
of typical star forming regions fails
to reproduce the strength of the high-J CO transitions observed
by \cite{bertoldi03b}. It is possible
to increase the gas heating in the CO-emitting region by increasing the
cosmic ray ionization rate. Such a high ionization rate can
mimic the higher gas heating due to an
increased X-ray flux associated with the powerful QSO hosted by this system.
We find a good fit to the observations with a gas density $n=\rm 10^5~cm^{-3}$,
a FUV field $\rm G_0=10^{3.8}$ times stronger than the average value
in the Galactic ISM, C/O=1/6, and a
cosmic ray ionization rate $\zeta$ per H nucleus
of $\rm \sim 10^{-16}~s^{-1}$, i.e. about 10 times
the Galactic value. The cosmic background radiation at $z=6.4$
has a temperature of about 22~K, much lower than the gas temperature
observed in this system, it may however slightly affect the lower CO
transitions, but the latter are less relevant for our model (especially
CO(1--0) which only has an upper limit).
The resulting model reproduces all the observed line
ratios and line-to-FIR ratios within a factor of 30\%.
This result shows that the C-poor scenario is consistent with the data.
On the other hand it is not a proof of a low C--to--O ratio.
Indeed, even using the C/O ratio typical of the local ISM the
results of the model are still in fair agreement with the data,
as shown in the third column of Tab.~\ref{mod}.
Summarizing, PDR models
with high densities and enhanced X-ray (cosmic ray) flux
can reproduce the observed properties of J1148+5251, both with
local ISM abundances and with a reduced C/O ratio predicted by
chemical evolutionary models. Tackling the issue of the chemical
abundances in detail will require the observation of additional lines.

\begin{table}[t!]
\caption{PDR model results
 ($n=10^5~\rm cm^{-3}$, $G_0=10^{3.8}$,
  $\zeta=1.8\times 10^{-16}~\rm s^{-1}$).}
\label{mod}     
\begin{tabular}{llll}
\hline
\hline
Ratio&Observed&\multicolumn{2}{c}{Model$^a$}\\
 \cline{3-4}
     &        & C/O=1/6  & C/O=local \\
\hline
$\rm L_{[CII]}/L_{FIR}$ &   $2.0 (\pm 0.9) \times10^{-4}$&  $1.4\times 10^{-4}$ & $3.8\times 10^{-4}$\\
$\rm L_{CO(6-5)}/L_{[CII]}$   & $6.6 (\pm 2.0) \times 10^{-2}$&   $5.7\times10^{-2}$  & $6.7\times 10^{-2}$\\
$\rm L_{CO(7-6)}/L_{CO(6-5)}$ &    $1.02 (\pm 0.17)$&                $0.85$  & $1.18$\\
$\rm L_{CO(3-2)}/L_{CO(6-5)}$ &    $0.22 (\pm 0.03)$&                $0.27$  & $0.17$ \\
$\rm L_{CO(1-0)}/L_{CO(6-5)}$ &  $<2.4\times 10^{-2}$&  $8.3\times 10^{-3}$& $4.4\times 10^{-3}$\\
\hline
\end{tabular}
Note: $^a$ The abundances for the models with ``local'' ISM fractions are
specifically O/H=3.2~10$^{-4}$ and C/H=1.4~10$^{-4}$ \citep[][]{savage96}, while
for the ``reduced'' C/O model C/H=5.6~10$^{-5}$ and O/H has been
kept to the same value as ``local''.
\end{table}

As discussed in the introduction, the [CII] line is produced by
the UV field in star forming regions and therefore can be used
as an indicator of the star formation rate (SFR).
\cite{boselli02} calibrate the luminosity
of [CII] in terms of SFR, but their calibration
only applies to low luminosity systems ($\rm L_{FIR}<10^{10.5}~L_{\odot}$).
We have attempted to derive the $\rm L_{[CII]}-SFR$ relation at high
luminosities ($  L_{FIR}>10^{12}L_{\odot}$).
One possibility is to combine
the relationship between far-IR luminosity and
star formation rate derived by \cite{kennicutt98} with the
$\rm L_{[CII]}/L_{FIR}$ ratio predicted by the model (where we use
the intermediate value of the $\rm L_{[CII]}/L_{FIR}$ obtained from the
two models in Tab.2.),
yielding the relation:
\begin{equation}
\rm SFR(M_{\odot}~yr^{-1}) \approx 6.5~10^{-7}
    \left( \frac{L_{[CII]}}{L_{\odot}} \right)
\end{equation}
From Eq.~1 and from the [CII] luminosity we infer a star formation rate in the host
of J1148+5251 of $\rm SFR \sim 3000 ~M_{\odot}~yr^{-1}$, consistent with
the SFR estimated from $\rm L_{FIR}$ \citep[][]{bertoldi03a}.
Therefore our result
demonstrates that the host galaxy of this quasar is the site of extremely powerful
massive star formation, and it is probably tracing the earliest stages of the 
black hole--spheroid co-evolution.

The first detection of the fine-structure line of C$^+$ in a
high redshifted object shows the potential use of this
emission line to investigate star formation and the physics
of the interstellar medium in the early universe.
Further detections \textit{and} imaging of C$^+$ in high-$z$ objects will
continued to be made by using existing instruments  and forthcoming
facilities, in particular with the Atacama
Large Millimeter Array (ALMA) for which this
topic is one of the primary scientific goals.

\begin{acknowledgements}
  We thank X. Tielens and P. Vanden Bout for useful comments.
  M.J.K.'s participation was made 
possible through a sabbatical leave from San Jose State University
and the support and hospitality of Arcetri Observatory.
   We thank the IRAM staff for their support during the observations.
   IRAM is supported by INSU/CNRS (France), MPG (Germany) and IGN (Spain).
This work has benefited from research funding from the European
Community's Sixth Framework Programme. R.M. acknowledges financial support from
MIUR under grant PRIN-03-02-23.

\end{acknowledgements}

\end{document}